\documentclass[aps,prc,superscriptaddress,twoside,twocolumn,nofootinbib,showpacs,floatfix]{revtex4-1}
\usepackage{amsmath,amssymb}
\usepackage{graphicx}
\usepackage{subfigure}
\usepackage[colorlinks]{hyperref}
\usepackage[usenames,dvipsnames]{color}
\pdfpagewidth=\paperwidth
\pdfpageheight=\paperheight
\allowdisplaybreaks
\newcommand*{\slashed}[1]{{#1\!\!\!/}}
\newcommand*{\hc}{\text{H.\,c.}}
\usepackage{newtxtext, newtxmath}
\usepackage{braket}

\begin{document}

\title{\boldmath Effects of $\Delta(1905)5/2^+$ on $K^*\Sigma$ photoproduction}

\author{Ai-Chao Wang}
\affiliation{School of Nuclear Science and Technology, University of Chinese Academy of Sciences, Beijing 100049, China}

\author{Wen-Ling Wang}
\affiliation{School of Physics and Nuclear Energy Engineering, Beihang University, Beijing 100191, China}

\author{Fei Huang}
\email[Corresponding author. Email: ]{huangfei@ucas.ac.cn}
\affiliation{School of Nuclear Science and Technology, University of Chinese Academy of Sciences, Beijing 100049, China}
\affiliation{Institute of High Energy Physics, Chinese Academy of Sciences, Beijing 100049, China}

\date{\today}

\begin{abstract}
The two-channel photoproductions of $\gamma p \to K^{*+} \Sigma^{0}$ and $\gamma p \to K^{*0} \Sigma^{+}$ are investigated based on an effective Lagrangian approach at the tree-level Born approximation. In addition to the $t$-channel $K$, $\kappa$, $K^*$ exchanges, the $s$-channel nucleon ($N$) and $\Delta$ exchanges, the $u$-channel $\Lambda$, $\Sigma$, $\Sigma^*$ exchanges, and the generalized contact term, we try to take into account the minimum number of baryon resonances in constructing the reaction amplitudes to describe the experimental data. It is found that by including the $\Delta(1905)5/2^+$ resonance with its mass, width, and helicity amplitudes taken from the Review of Particle Physics [Particle Data Group, C. Patrignani {\it et al.}, Chin. Phys. C {\bf 40}, 100001 (2016)], the calculated differential and total cross sections for these two reactions are in good agreement with the experimental data. An analysis of the reaction mechanisms shows that the cross sections of $\gamma p \to K^{*+}\Sigma^{0}$ are dominated by the $s$-channel $\Delta(1905)5/2^+$ exchange at low energies and $t$-channel $K^*$ exchange at high energies, with the $s$-channel $\Delta$ exchange providing significant contributions in the near-threshold region. For $\gamma p \to K^{*0}\Sigma^{+}$, the angular dependences are  dominated by the $t$-channel $K$ exchange at forward angles and the $u$-channel $\Sigma^*$ exchange at backward angles, with the $s$-channel $\Delta$ and $\Delta(1905)5/2^+$ exchanges making considerable contributions at low energies. Predictions are given for the beam, target, and recoil asymmetries for both reactions.
\end{abstract}

\pacs{25.20.Lj, 13.60.Le, 14.20.Gk}

\keywords{$\Sigma K^*$ photoproduction, effective Lagrangian approach, nucleon resonances}

\maketitle

\section{Introduction}   \label{Sec:intro}

The study of nucleon resonances ($N^*$'s) has always been of great interest in hadron physics, with one
of the reasons being that the structures, parameters, and microscopic production mechanisms of $N^*$'s  are essential for our understanding of the nonperturbative behavior of quantum chromodynamics (QCD), the theory for strong interactions.  Currently, most of our knowledge of $N^*$'s is coming from the $\pi N$ scattering and $\pi$ photoproduction reactions. Since $K^* \Sigma$ has a much higher threshold than $\pi N$, its photoproduction off nucleon is more suitable to investigate the $N^*$'s in a less explored higher $N^*$ mass region.

Experimentally, the $K^* \Sigma$ photoproduction process has been investigated by
several collaborations \cite{Hleiqawi:2005sz,Hleiqawi:2007ad,
Nanova:2008kr,Hwang2012,Wei:2013}. The most recent data for $\gamma p \to K^{*+} \Sigma^0$ have been reported by the CLAS Collaboration at the Thomas Jefferson National Accelerator Facility (JLab) in 2013 \cite{Wei:2013}, where the first high-statistics differential cross-section and total cross-section data were presented at center-of-mass energies from threshold up to $2.8$ GeV. For the other reaction $\gamma p \to K^{*0} \Sigma^+$, so far the differential cross-section data are available from both the CLAS Collaboration \cite{Hleiqawi:2007ad} and the CBELSA/TAPS Collaboration \cite{Nanova:2008kr}. In this work, since we are seeking for a combined analysis of the data for both $\gamma p \to K^{*+} \Sigma^0$ and $\gamma p \to K^{*0} \Sigma^+$ reactions, we will concentrate on CLAS's data to keep the consistency for the data.

Theoretically, several works have already been devoted to the study of photoproductions of $K^{*+} \Sigma^0$ and $K^{*0} \Sigma^+$ off nucleon, employing a chiral quark model  \cite{Zhao:2001jw} or an effective Lagrangian approach \cite{Oh:2006in,Kim:2013,Kim:20132}. It is pointed out in Ref.~\cite{Oh:2006in} based on a study of the $K^{*0} \Sigma^+$ photoproduction that the $t$-channel $\kappa$-meson exchange may contribute significantly to $K^{*} \Sigma$ photoproduction, rather different from the process $\gamma p \to K^{*+} \Lambda$ where the dominate $t$-channel contribution is found to be the $K$-meson exchange \cite{Kim:2014,Wang:2017}. References \cite{Kim:2013,Kim:20132} provide so far the only theoretical analysis of the first high-statistics differential and total cross-section data for $\gamma p \to K^{*+}\Sigma^0$ reported by the CLAS Collaboration in 2013 \cite{Wei:2013}. Note that the theoretical results of Ref.~\cite{Kim:2013} and Ref.~\cite{Kim:20132} are the same, but a comparison with the data was only given in the latter reference. In Refs.~\cite{Kim:2013,Kim:20132}, the resonances $N(2080)3/2^-$, $N(2090)1/2^-$, $N(2190)7/2^-$, $N(2200)5/2^-$, $\Delta(2150)1/2^-$, $\Delta(2200)7/2^-$, and $\Delta(2390)7/2^+$ have been introduced in addition to the $s$-channel $N$ and $\Delta$ exchanges and $t$- and $u$-channel interactions to describe the data. It is claimed that the resonance contributions gave only negligible effects while the contributions of $t$-channel $K$ exchange and the $s$-channel $\Delta$ exchange are crucial for both $\gamma p \to K^{*+} \Sigma^0$ and $\gamma p \to K^{*0} \Sigma^+$  reaction processes.

\begin{figure*}[tbp]
\includegraphics[width=0.8\textwidth]{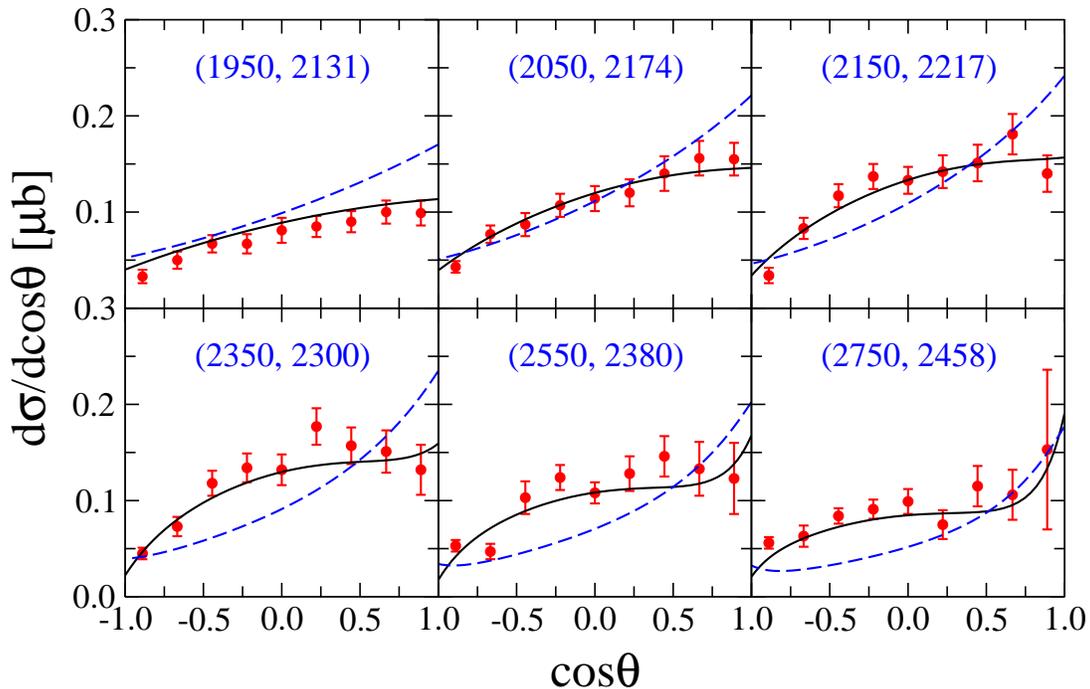}
\caption{Status of theoretical description of the differential
cross sections for $\gamma p\to K^{*+}\Sigma^0$ at selected energies.
The numbers in parentheses denote the photon laboratory
incident energy (left number) and the total center-of-mass energy of the system
(right number). The blue dashed lines represent the results from
Ref.~\cite{Kim:20132}, and the black solid lines denote our theoretical results.
 The scattered symbols are the most recent data from CLAS Collaboration \cite{Wei:2013}.}
\label{Fig:status}
\end{figure*}

Although the theoretical results from Refs.~\cite{Kim:2013,Kim:20132} are in qualitative agreement with the CLAS data, there are still some room for improvements in their results for the differential cross sections for $\gamma p \to K^{*+} \Sigma^0$. Figure \ref{Fig:status} illustrates this issue; there, a comparison of the differential cross sections from the theoretical calculation of Ref.~\cite{Kim:20132} (blue dashed lines) with the most recent CLAS data \cite{Wei:2013} (scattered symbols) at some selected energies is shown. The numbers in parentheses denote the photon laboratory incident energy, $E_\gamma$ (left number), and the total center-of-mass energy of the system, $W$ (right number). The black solid lines represent the results from our present work, which will be discussed in detail in Sec.~\ref{Sec:results}. It is clearly seen from Fig.~\ref{Fig:status} that there is still some room for improvement in the differential cross-section results of Ref.~\cite{Kim:20132}.

In this work, we perform a two-channel combined analysis of $\gamma p \to K^{*+} \Sigma^{0}$ and $\gamma p \to K^{*0} \Sigma^{+}$ reactions based on an effective Lagrangian approach in the tree-level approximation. We believe that an analysis of the CLAS data for $K^*\Sigma$ photoproduction reactions independent of Refs.~\cite{Kim:2013,Kim:20132} is necessary and meaningful. Moreover, we expect that a better description of the data for these two reactions will allow for a more reliable understanding of the reaction mechanisms and the associated resonance contents and parameters. Unlike Refs.~\cite{Kim:2013,Kim:20132} where seven resonances, namely $N(2080)3/2^-$, $N(2090)1/2^-$, $N(2190)7/2^-$, $N(2200)5/2^-$, $\Delta(2150)1/2^-$, $\Delta(2200)7/2^-$, and $\Delta(2390)7/2^+$, have been considered in addition to the $N$ and $\Delta$ exchanges, the strategy adopted in the present work in choosing the nucleon resonances is that we introduce the nucleon resonances as few as possible to reproduce the data. We find that if we only consider the contributions from the $t$-channel $K$, $\kappa$, $K^*$ exchanges, the $s$-channel $N$, $\Delta$ exchanges, the $u$-channel $\Lambda$, $\Sigma$, $\Sigma^*$ exchanges, and the generalized contact current, the fitting quality of the CLAS high-statistics differential and total cross-section data for $\gamma p \to K^{*+} \Sigma^{0}$ and $\gamma p \to K^{*0} \Sigma^{+}$ will be worse than that illustrated by the dashed lines of Fig.~\ref{Fig:status}, which in our opinion can not be treated as an acceptable description of the data. We then introduce one resonance in constructing the reaction amplitudes. We check one by one the near-threshold four-star or three-star resonances advocated in the 2016 edition of Review of Particle Physics (RPP) \cite{Patrignani:2016}, with the resonance mass, width, and helicity amplitudes being fixed to be the averaged values reported in RPP \cite{Patrignani:2016}. It is found that the data can be well described by including the $\Delta(1905)5/2^+$ resonance, which is rated as a four-star resonance in the 2016 edition of RPP \cite{Patrignani:2016}. An analysis of the reaction mechanisms shows that the cross sections of $\gamma p \to K^{*+}\Sigma^{0}$ are dominated by the $s$-channel $\Delta(1905)5/2^+$ exchange at low energies and $t$-channel $K^*$ exchange at high energies. The $s$-channel $\Delta$ exchange also provides significant contributions to this reaction in the near-threshold region. For $\gamma p \to K^{*0}\Sigma^{+}$, the angular dependences are  dominated by the $t$-channel $K$ exchange at forward angles and the $u$-channel $\Sigma^*$ exchange at backward angles, with the $s$-channel $\Delta$ and $\Delta(1905)5/2^+$ exchanges making considerable contributions at low energies. We also present our predictions for the beam, target, and recoil asymmetries for these two reactions for future experiments. 

Introducing another three-star or four-star resonance instead of $\Delta(1905)5/2^+$ in constructing the reaction amplitudes will result in a much larger $\chi^2$ and obvious discrepancies in comparison with the data. We do not attempt to include the one-star or two-star resonances whose masses, widths, and helicity amplitudes are not well determined in RPP \cite{Patrignani:2016} or introduce one more resonance in constructing the reaction amplitudes, as doing so will lead to much more adjustable parameters that cannot be well constrained by the cross-section data alone, which are so far the only data we have.

The present paper is organized as follows. In Sec.~\ref{Sec:formalism}, we briefly introduce the framework of our theoretical model, including the generalized contact current, the effective interaction Lagrangians, the resonance propagators and the phenomenological form factors employed in the present work. In Sec.~\ref{Sec:results}, we present our theoretical results, and a discussion of the contributions of various individual terms in each reaction is given as well. Furthermore, the spin observables including the beam, target, and recoil asymmetries are also shown and discussed in this section. Finally, a brief summary and conclusions are given in Sec.~\ref{sec:summary}.

\section{Formalism}  \label{Sec:formalism}

\begin{figure}[tbp]
\centering
{\vglue 0.15cm}
\subfigure[~$s$ channel]{
\includegraphics[width=0.45\columnwidth]{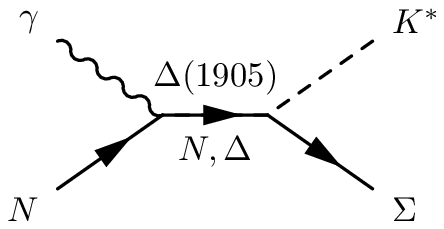}}  {\hglue 0.4cm}
\subfigure[~$t$ channel]{
\includegraphics[width=0.45\columnwidth]{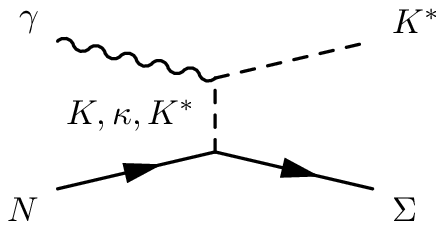}} \\[6pt]
\subfigure[~$u$ channel]{
\includegraphics[width=0.45\columnwidth]{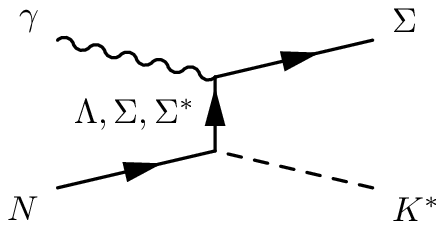}} {\hglue 0.4cm}
\subfigure[~Interaction current]{
\includegraphics[width=0.45\columnwidth]{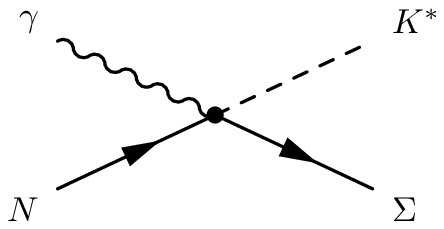}}
\caption{Generic structure of the $K^*$ photoproduction amplitude for $\gamma N\to K^{*}\Sigma$. Time proceeds from left to right.}
\label{FIG:feymans}
\end{figure}

Following a full field theoretical approach of Refs.~\cite{Haberzettl:1997,Haberzettl:2006,Huang:2012,Huang:2013}, the full reaction amplitude for $\gamma N \to K^* \Sigma$ can be expressed as
\begin{eqnarray}
M^{\nu\mu} = M^{\nu\mu}_s + M^{\nu\mu}_t + M^{\nu\mu}_u + M^{\nu\mu}_{\rm int},  \label{eq:amplitude}
\end{eqnarray}
with $\nu$ and $\mu$ being the Lorentz indices of vector meson $K^*$ and photon $\gamma$, respectively. The first three terms $M^{\nu\mu}_s$, $M^{\nu\mu}_t$, and $M^{\nu\mu}_u$ stand for the $s$-, $t$-, and $u$-channel pole diagrams, respectively, with $s$, $t$, and $u$ being the Mandelstam variables of the internally exchanged particles. They arise from the photon attaching to the external particles  in the underlying $\Sigma NK^*$ interaction vertex. The last term, $M^{\nu\mu}_{\rm int}$, stands for the interaction current, which arises from the photon attaching to the internal structure of the $\Sigma NK^*$ interaction vertex. All four terms in Eq.~(\ref{eq:amplitude}) are diagrammatically depicted in Fig.~\ref{FIG:feymans}.

In the present work, the following contributions, as shown in Fig.~\ref{FIG:feymans}, are considered in constructing the $s$-, $t$-, and $u$-channel amplitudes: (i) $N$, $\Delta$ and $\Delta(1905)5/2^+$ exchanges in the $s$ channel, (ii) $K$, $\kappa$, and $K^*$ meson exchanges in the $t$ channel, and (iii) $\Lambda$, $\Sigma$, and $\Sigma^*(1385)$ hyperon exchanges in the $u$ channel. The exchanges of other hyperon resonances with higher masses in the $u$ channel are tested to have tiny contributions and thus are omitted in the present work in order to reduce the model parameters. Using an effective Lagrangian approach, one can, in principle, obtain explicit expressions for these amplitudes. However, the exact calculation of the interaction current $M^{\nu\mu}_{\rm int}$ is impractical, as it obeys a highly non-linear equation and contains diagrams with very complicated interaction dynamics. Furthermore, the introduction of phenomenological form factors makes it impossible to calculate the interaction current exactly even in principle. Following Refs.~\cite{Haberzettl:1997,Haberzettl:2006,Huang:2012,Huang:2013}, we model the interaction current by a generalized contact current, that accounts effectively for the interaction current arising from the unknown parts of the underlying microscopic model,
\begin{eqnarray}
M^{\nu\mu}_{\rm int} = \Gamma^\nu_{\Sigma N K^*}(q) C^\mu + M_{\rm KR}^{\nu\mu} f_t.  \label{eq:Mint}
\end{eqnarray}
Here $\nu$ and $\mu$ are Lorentz indices for $K^*$ and $\gamma$, respectively; $\Gamma^\nu_{\Sigma N K^*}(q)$ is the vertex function of $\Sigma N K^*$ coupling given by the Lagrangian of Eq.~(\ref{eq:L_SNKst}),
\begin{eqnarray}
\Gamma^\nu_{\Sigma N K^*}(q) = - i g_{\Sigma N K^*}\left[\gamma^{\nu}-i\frac{\kappa_{\Sigma N K^*}}{2M_N}\sigma^{\nu \alpha}q_{\alpha}\right],
\end{eqnarray}
with $q$ being the four-momentum of the outgoing $K^*$ meson; $M_{\rm KR}^{\nu\mu}$ is the Kroll-Ruderman term given by the Lagrangian of Eq.~(\ref{eq:L_gLNKst}),
\begin{eqnarray}
M_{\rm KR}^{\nu \mu}= g_{\Sigma N K^*}\frac{\kappa_{\Sigma N K^*}}{2M_N}\sigma^{\nu\mu} Q_{K^*},
\end{eqnarray}
with $Q_{K^*}$ being the electric charge of $K^*$; $f_t$ is the phenomenological form factor attached to the amplitude of $t$-channel $K^*$ exchange, which is given in Eq.~(\ref{eq:ff_M}); $C^\mu$ is an auxiliary current, which is non-singular, introduced to ensure that the full photoproduction amplitude of Eq.~(\ref{eq:amplitude}) satisfies the generalized WTI and thus is fully gauge invariant. Following Refs.~\cite{Haberzettl:2006,Huang:2012}, we choose $C^\mu$ for $\gamma p \to K^{*+} \Sigma^0$ as
\begin{equation}
C^\mu =  - Q_{K^*} \frac{f_t-\hat{F}}{t-q^2}  (2q-k)^\mu - Q_N \frac{f_s-\hat{F}}{s-p^2} (2p+k)^\mu,
\end{equation}
with
\begin{equation} \label{eq:Fhat-Kstp}
\hat{F} = 1 - \hat{h} \left(1 -  f_s\right) \left(1 - f_t\right),
\end{equation}
and for $\gamma p \to K^{*0} \Sigma^+$ as
\begin{equation}
C^\mu =  - Q_{\Sigma} \frac{f_{\mu}-\hat{F}}{u-p'^{2}}  (2 p'-k)^\mu - Q_N \frac{f_s-\hat{F}}{s-p^2} (2p+k)^\mu,
\end{equation}
with
\begin{equation} \label{eq:Fhat-Kst0}
\hat{F} = 1 - \hat{h} \left(1 -  f_u\right) \left(1 - f_s\right).
\end{equation}
Here $p$, $p'$, $q$, and $k$ are four-momenta for incoming $N$, outgoing $\Sigma$, outgoing $K^*$, and incoming photon, respectively; $Q_{N\left(K^*,\Sigma\right)}$ is the electric charge of $N\left(K^*,\Sigma\right)$; $f_s$ is the phenomenological form factor for $s$-channel $N$ exchange, $f_t$ for $t$-channel $K^*$ exchange, and  $f_u$ for $u$-channel $\Sigma$ exchange, respectively. $\hat{h}$ is an arbitrary function, except that it should go to unity in the high-energy limit to prevent the ``violation of scaling behavior'' \cite{Drell:1972}. For the sake of simplicity, in the present work it is taken to be $\hat{h}=1$.

In the rest of this section,  we present the effective Lagrangians, the resonance propagators, and the phenomenological form factors employed in the present work.

\subsection{Effective Lagrangians} \label{Sec:Lagrangians}

The effective interaction Lagrangians used in the present work for the production amplitudes are given below. For further convenience, we define the operators
\begin{equation}
\Gamma^{(+)}=\gamma_5  \quad  \text{and} \quad  \Gamma^{(-)}=1,
\end{equation}
and the field-strength tensors
\begin{eqnarray}
{K^*}^{\mu\nu} &=& \partial^{\mu} {K^*}^{\nu} - \partial^{\nu} {K^*}^{\mu},  \\[6pt]
F^{\mu\nu} &=& \partial^{\mu}A^\nu-\partial^{\nu}A^\mu,
\end{eqnarray}
with ${K^{*\mu}}$ and $A^\mu$ denoting the $K^*$ vector-meson field and electromagnetic field, respectively.

The electromagnetic interaction Lagrangians required to calculate the non-resonant Feynman diagrams are
\begin{eqnarray}
{\cal L}_{NN\gamma} &=& -\,e \bar{N} \left[ \left( \hat{e} \gamma^\mu - \frac{ \hat{\kappa}_N} {2M_N}\sigma^{\mu \nu}\partial_\nu\right) A_\mu\right] N, \\[6pt]
{\cal L}_{\gamma K^* K^* } &=& -\,e \left({K^*}^{\nu} \times K^*_{\mu\nu}\right)_3 A^\mu, \\[6pt]
{\cal L}_{\gamma \kappa{K^*}} &=& e\frac{g_{\gamma \kappa{K^*}}}{2M_{K^*}}F^{\mu \nu}K^*_{\mu \nu}\kappa, \label{Lag:gkaKst}    \\[6pt]
{\cal L}_{\gamma K{K^*}} &=& e\frac{g_{\gamma K{K^*}}}{M_K}\varepsilon^{\alpha \mu \lambda \nu}\left(\partial_\alpha A_\mu\right)\left(\partial_\lambda K\right)K^*_\nu, \label{Lag:gKKst} \\[6pt]
{\cal L}_{\Sigma \Sigma \gamma} &=& -\,e \bar{\Sigma}\left[\left(\hat{e}\gamma^{\mu} - \frac{\hat{\kappa}_{\Sigma}}{2M_N}\sigma^{\mu\nu}\partial_\nu \right)A_{\mu}\right]\Sigma,   \\[6pt]
{\cal L}_{\Sigma \Lambda \gamma} &=& e\frac{\kappa_{\Sigma \Lambda}}{2M_N}\bar{\Lambda}\sigma^{\mu \nu}\left(\partial_\nu A_\mu\right)\Sigma^0 + \hc,  \\[6pt]
{\cal L}_{\Sigma^* \Sigma \gamma} &=& ie\frac{g^{(1)}_{\Sigma^* \Sigma \gamma}}{2M_N}\bar{\Sigma}\gamma_\nu \gamma_5 F^{\mu \nu} \Sigma^*_\mu \nonumber \\
&& -\,e\frac{g^{(2)}_{\Sigma^* \Sigma \gamma}}{\left(2M_N\right)^2} \left(\partial_\nu\bar{\Sigma}\right) \gamma_5 F^{\mu \nu}{\Sigma}^*_\mu + \hc,\\[6pt]
{\cal L}_{\Delta N \gamma} &=& - ie\frac{g^{(1)}_{\Delta N \gamma}}{2M_N}\bar{\Delta}_{\mu}\gamma_\nu \gamma_5 F^{\mu \nu} N \nonumber \\
&& +\,e\frac{g^{(2)}_{\Delta N \gamma}}{\left(2M_N\right)^2} \left(\bar{\Delta}{_\mu}\right) \gamma_5 F^{\mu \nu}\partial_\nu N + \hc,
\end{eqnarray}
where $e$ is the elementary charge unit and $\hat{e}$ stands for the charge operator; $\hat{\kappa}_N = \kappa_p\left(1+\tau_3\right)/2 + \kappa_n\left(1-\tau_3\right)/2$, with the anomalous magnetic moments $\kappa_p=1.793$ and $\kappa_n=-1.913$;
$\hat{\kappa}_\Sigma = \kappa_{\Sigma^+}\left( 1 + \hat{e}\right)/2 + \kappa_{\Sigma^-}\left(1-\hat{e}\right)/2$ with the anomalous magnetic moment $\kappa_{\Sigma^+} = 1.458$ and $\kappa_{\Sigma^-}= -0.16$; $\kappa_{\Sigma\Lambda}=-1.61$ is the anomalous magnetic moment for $\Sigma^0\to \Lambda \gamma$ transition; $M_N$, $M_K$, and $M_{K^*}$ stand for the masses of $N$, $K$, and $K^*$, respectively; $\varepsilon^{\alpha \mu \lambda \nu}$ is the totally antisymmetric Levi-Civita tensor with $\varepsilon^{0123}=1$. The coupling constants $g_{\gamma\kappa^\pm K^{*\pm}}=0.214$ and $g_{\gamma\kappa^0 K^{*0}}=-2g_{\gamma\kappa^\pm K^{*\pm}}$ are taken from Refs.~\cite{Kim:2011,Kim:2014}, determined by a vector-meson dominance model proposed by Black {\it et al.} \cite{BHS02}. The value of the electromagnetic coupling $g_{\gamma K K^*}$ is determined by fitting the radiative decay width of $K^*\to K\gamma$ given by the RPP \cite{Patrignani:2016}, which leads to $g_{\gamma K^\pm K^{*\pm}}=0.413$ and $g_{\gamma K^0 K^{*0}}=-0.631$ with the sign inferred from $g_{\gamma \pi \rho}$ \cite{Garcilazo:1993} via the flavor SU(3) symmetry considerations in conjunction with the vector-meson dominance assumption. The coupling constants $g^{(1)}_{\Sigma^{*+} \Sigma^{+} \gamma}$ and $g^{(2)}_{\Sigma^{*+} \Sigma^{+} \gamma}$ get one constraint from the RPP value of the partial decay width $\Gamma_{\Sigma^{*+}\to \Sigma^{+} \gamma}=0.252$ MeV \cite{Patrignani:2016}, therefore only one of them is free. In the present work we treat $g^{(2)}_{\Sigma^{*+} \Sigma^{+} \gamma}/g^{(1)}_{\Sigma^{*+} \Sigma^{+} \gamma}$ as a fitting parameter. The couplings $g^{(1)}_{\Sigma^{*0} \Sigma^{0} \gamma}$ and $g^{(2)}_{\Sigma^{*0} \Sigma^{0} \gamma}$ are both treated as fitting parameters, as we do not have any decay information for $\Sigma^{*0}\to\Sigma^0\gamma$ from RPP \cite{Patrignani:2016}. The $\Delta N\gamma$ couplings are determined by the RPP values of $\Delta\to N\gamma$ helicity amplitudes, which leads to $g^{(1)}_{\Delta N \gamma}=-4.18$ and $g^{(2)}_{\Delta N \gamma}=4.327$.

The effective Lagrangians for meson-baryon interactions are
\begin{eqnarray}
{\cal L}_{\Sigma N {K^*}} &=&  -\, g_{\Sigma N {K^*}} \bar{\Sigma} \left[\left(\gamma^\mu-\frac{\kappa_{\Sigma N {K^*}}}{2M_N}\sigma^{\mu \nu}\partial_\nu\right)K^*_\mu\right] N + \hc, \nonumber \\  \label{eq:L_SNKst}  \\[6pt]
{\cal L}_{\Lambda N {K^*}} &=&  -\, g_{\Lambda N {K^*}} \bar{\Lambda} \left[\left(\gamma^\mu-\frac{\kappa_{\Lambda N {K^*}}}{2M_N}\sigma^{\mu \nu}\partial_\nu\right)K^*_\mu\right] N + \hc,   \nonumber \\\label{eq:L_LNKst}  \\[6pt]
{\cal L}_{\Sigma N\kappa} &=& -\, g_{\Sigma N\kappa} \bar{\Sigma} \kappa N + \hc, \\[6pt]
{\cal L}_{\Sigma NK} &=& -\, g_{\Sigma NK}\bar{\Sigma}\Gamma^{(+)} \left[\left(i\lambda + \frac{1-\lambda}{2M_N} \slashed{\partial}\right)K\right] N + \hc,  \label{eq:L_LNK}    \\[6pt]
{\cal L}_{\Sigma^* N{K^*}} &=& -\, i\frac{g_{\Sigma^* N K^*}^{(1)}}{2M_{N}}{\bar\Sigma}^*_\mu \gamma_\nu \gamma_5 {K^*}^{\mu \nu}N \nonumber \\
&& +\, \frac{g_{\Sigma^* N K^*}^{(2)}}{\left(2M_{N}\right)^2}{\bar\Sigma}^*_\mu \gamma_5 {K^*}^{\mu \nu}\partial_\nu N \nonumber \\
&& -\, \frac{g_{\Sigma^* N K^*}^{(3)}}{\left(2M_{N}\right)^2}{\bar\Sigma}^*_\mu \gamma_5\left(\partial_\nu {K^{*\mu \nu}}\right) N + \hc. \\[6pt]
{\cal L}_{\Delta \Sigma {K^*}} &=& +\, i\frac{g_{\Delta \Sigma K^*}^{(1)}}{2M_N}{\bar \Sigma} \gamma_\nu \gamma_5 {K^*}^{\mu \nu}\Delta_\mu  \nonumber \\
&& -\, \frac{g_{\Delta \Sigma K^*}^{(2)}}{\left(2M_{N}\right)^2}\partial_\nu{\bar \Sigma} \gamma_5 {K^*}^{\mu \nu} \Delta_\mu \nonumber \\
&& +\, \frac{g_{\Delta \Sigma K^*}^{(3)}}{\left(2M_{N}\right)^2}{\bar\Sigma} \gamma_5\left(\partial_\nu {K^{*\mu \nu}}\right) \Delta_\mu + \hc,
\end{eqnarray}
where the parameter $\lambda$ was introduced in ${\cal L}_{\Sigma NK}$ to interpolate between the pseudovector $(\lambda=0)$ and the pseudoscalar $(\lambda=1)$ couplings. Following our previous work on $\gamma p \to K^{*+}\Lambda$ where $\lambda=1$ is chosen for ${\cal L}_{\Lambda NK}$ \cite{Wang:2017}, we choose $\lambda=1$ for ${\cal L}_{\Sigma NK}$ based on the SU(3) flavor symmetry. The coupling constants $g_{\Sigma NK}$, $g_{\Lambda N K^*}$, $\kappa_{\Lambda N K^*}$ and $g^{(1)}_{\Sigma^* NK^*}$ are fixed by the flavor SU(3) symmetry \cite{Swart:1963,Ronchen:2013},
\begin{eqnarray}
g_{\Sigma NK} &=& \frac{1}{5}  g_{NN\pi} = 2.692, \\[6pt]
g_{\Lambda N K^*} &=&  -\, \frac{1}{2\sqrt{3}} g_{NN\omega} - \frac{\sqrt{3}}{2} g_{NN\rho} = -6.21,  \\[6pt]
\kappa_{\Lambda N K^*} &=&  \frac{f_{\Lambda N K^*}}{g_{\Lambda N K^*}}  = -\frac{\sqrt{3}}{2} \frac{f_{NN\rho}}{g_{\Lambda N K^*}} = 2.76,    \\[6pt]
g^{(1)}_{\Sigma^* NK^*} &=& -\,\frac{1}{\sqrt{6}} g_{\Delta N\rho} = 15.96,\\[6pt]
\nonumber
\end{eqnarray}
where the empirical values $g_{NN\pi}=13.46$, $g_{NN\rho}=3.25$, $g_{NN\omega}=11.76$, $\kappa_{NN\rho}=f_{NN\rho}/g_{NN\rho}=6.1$, and $g_{\Delta N \rho } = -39.1$ from Refs.~\cite{Huang:2012,Ronchen:2013}
are quoted. As the $g^{(2)}$ and $g^{(3)}$ terms in the $\Delta N\rho$ interactions have never been seriously studied in literature, the corresponding couplings for the $\Sigma^* NK^*$ interactions, i.e., $g_{\Sigma^* N K^*}^{(2)}$ and $g_{\Sigma^* N K^*}^{(3)}$, cannot be determined via flavor SU(3) symmetry, and we ignore these two terms in the present work, following Refs.~\cite{Kim:2011,Kim:2014,Wang:2017}. The $\Delta\Sigma K^*$ couplings are found to be sensitive to the fitting quality of the cross-section results, we thus leave the coupling $g_{\Delta \Sigma K^*}^{(1)}$ as a fit parameter but ignore the $g_{\Delta \Sigma K^*}^{(2)}$ and $g_{\Delta \Sigma K^*}^{(3)}$ terms. Following Refs.~\cite{Kim:2013,Kim:20132}, the coupling constants $g_{\Sigma NK^*}=-2.46$,  $\kappa_{\Sigma N K^*}=-0.47$, $g_{\Sigma N\kappa}=-5.32$ are taken from Nijmegen model (NSC97a) \cite{Stoks:1999}, determined by a fit to the $\Lambda N-\Sigma N$ scattering data. 

The resonance $\Delta(1905)5/2^+$ electromagnetic and hadronic coupling Lagrangians are
\begin{eqnarray}
{\cal L}_{RN\gamma}^{5/2+} & = & e\frac{g_{RN\gamma}^{(1)}}{\left(2M_N\right)^2}\bar{R}_{\mu \alpha}\gamma_\nu \left(\partial^{\alpha} F^{\mu \nu}\right)N \nonumber \\
&& +\, ie\frac{g_{RN\gamma}^{(2)}}{\left(2M_N\right)^3}\bar{R}_{\mu \alpha} \left(\partial^\alpha F^{\mu \nu}\right)\partial_\nu N +  \hc,  \\[6pt]
{\cal L}_{R\Sigma {K^*}}^{5/2+} &= & \frac{g_{R\Sigma {K^*}}^{(1)}}{\left(2M_N\right)^2}\bar{R}_{\mu \alpha}\gamma_\nu \left(\partial^{\alpha} K^{* \mu \nu}\right) \Sigma   \nonumber \\
&& +\,  i\frac{g_{R\Sigma K^*}^{(2)}}{\left(2M_N\right)^3}\bar{R}_{\mu \alpha} \left(\partial^\alpha {K^*}^{\mu \nu}\right)\partial_\nu \Sigma  \nonumber \\
&& -\,  i\frac{g_{R\Sigma {K^*}}^{(3)}}{\left(2M_N\right)^3}\bar{R}_{\mu \alpha}  \left(\partial^\alpha \partial_\nu {K^*}^{\mu \nu}\right) \Sigma + \hc,
\end{eqnarray}
where $R$ designates the $\Delta(1905)5/2^+$ resonance, and the superscripts of ${\cal L}_{RN\gamma}$ and ${\cal L}_{R\Sigma K^*}$ denote the spin and parity of the resonance $R$. The electromagnetic couplings for $\Delta(1905)5/2^+$ are fixed by use of the RPP values of helicity amplitudes for $\Delta(1905)5/2^+\to N\gamma$ \cite{Patrignani:2016}, while the hadronic couplings $g_{R\Sigma K^*}^{(i)}$ $(i=1,2,3)$ are treated as fit parameters.

The effective Lagrangian for the Kroll-Ruderman term of $\gamma N\to \Sigma K^*$ reads
\begin{equation}
{\cal L}_{\gamma N \Sigma {K^*}} =  -\, i  g_{\Sigma N {K^*}} \frac{\kappa_{\Sigma N {K^*}}}{2M_N} \bar{\Sigma} \sigma^{\mu \nu} A_\nu  \hat{Q}_{K^*} K^*_\mu N + \hc,   \label{eq:L_gLNKst}
\end{equation}
with $\hat{Q}_{K^*}$ being the electric charge operator of the outgoing $K^*$ meson. This interaction Lagrangian is obtained by the minimal gauge substitution $\partial_\mu \to {\cal D}_\mu\equiv \partial_\mu-i\hat{Q}_{K^*}A_\mu$ in the $\Sigma N K^*$ interaction Lagrangian of Eq.~(\ref{eq:L_SNKst}).

\subsection{Resonance propagators}

Apart from $N$, we have $\Delta$ and $\Delta(1905)5/2^+$ resonances in $s$ channel.
Following Refs.~\cite{Behrends:1957,Fronsdal:1958,Zhu:1999}, the prescriptions of the propagators for resonances with spin-$3/2$, and -$5/2$ are
\begin{eqnarray}
S_{3/2}(p) &=&  \frac{i}{\slashed{p} - M_R + i \Gamma/2} \left( \tilde{g}_{\mu \nu} + \frac{1}{3} \tilde{\gamma}_\mu \tilde{\gamma}_\nu \right),  \\[6pt]
S_{5/2}(p) &=&  \frac{i}{\slashed{p} - M_R + i \Gamma/2} \,\bigg[ \, \frac{1}{2} \big(\tilde{g}_{\mu \alpha} \tilde{g}_{\nu \beta} + \tilde{g}_{\mu \beta} \tilde{g}_{\nu \alpha} \big)  \nonumber \\
&& -\, \frac{1}{5}\tilde{g}_{\mu \nu}\tilde{g}_{\alpha \beta}  + \frac{1}{10} \big(\tilde{g}_{\mu \alpha}\tilde{\gamma}_{\nu} \tilde{\gamma}_{\beta} + \tilde{g}_{\mu \beta}\tilde{\gamma}_{\nu} \tilde{\gamma}_{\alpha}  \nonumber \\
&& +\, \tilde{g}_{\nu \alpha}\tilde{\gamma}_{\mu} \tilde{\gamma}_{\beta} +\tilde{g}_{\nu \beta}\tilde{\gamma}_{\mu} \tilde{\gamma}_{\alpha} \big) \bigg],
\end{eqnarray}
where
\begin{eqnarray}
\tilde{g}_{\mu \nu} &=& -\, g_{\mu \nu} + \frac{p_{\mu} p_{\nu}}{M_R^2}, \\[6pt]
\tilde{\gamma}_{\mu} &=& \gamma^{\nu} \tilde{g}_{\nu \mu} = -\gamma_{\mu} + \frac{p_{\mu}\slashed{p}}{M_R^2}.
\end{eqnarray}

\subsection{Form factors}

Each hadronic vertex obtained from the Lagrangians given in Sec.~\ref{Sec:Lagrangians} is accompanied with a phenomenological form factor to parametrize the structure of the hadrons and to normalize the behavior of the production amplitude. Following Refs.~\cite{Kim:2011,Kim:2014,Wang:2017}, for intermediate baryon exchange we take the form factor as
\begin{eqnarray}
f_B(p^2) = \left(\frac{\Lambda_B^4}{\Lambda_B^4+\left(p^2-M_B^2\right)^2}\right)^n,  \label{eq:ff_B}
\end{eqnarray}
where $p$ denotes the four-momentum of the intermediate baryon, the exponent $n$ is taken to be $2$ for all baryon exchanges, and $M_B$ is the mass for exchanged baryon $B$. The cutoff $\Lambda_B$ is treated as a fitting parameter for each exchanged baryon, except for the $s$ channel, where a common cutoff $\Lambda_s$ is introduced for all $N$, $\Delta$, and $\Delta(1905)5/2^+$ exchanges. 
For intermediate meson exchange, we take the form factor as
\begin{eqnarray}
f_M(q^2) = \left(\frac{\Lambda_M^2-M_M^2}{\Lambda_M^2-q^2}\right)^m, \label{eq:ff_M}
\end{eqnarray}
where $q$ represents the four-momentum of the intermediate meson, the exponent $m$ is taken to be $2$  for all meson exchanges, and $M_M$ and $\Lambda_M$ designate the mass and cutoff mass of exchanged meson $M$. We choose $M_\kappa = 800$ MeV, and for $K$ and $K^*$ exchanges, the experimental values are used for their masses. As the results are tested to be not sensitive to the $\kappa$ exchange, we use the same cutoff $\Lambda_{K,\kappa}$ for $\kappa$ and $K$ exchanges.

Note that the gauge-invariance feature of our photoproduction amplitude is independent of the specific form of the form factors, which is different from Refs.~\cite{Kim:2013,Kim:20132} where a common form factor is introduced in the reaction amplitudes in order to preserve gauge invariance.

\section{Results and discussion}   \label{Sec:results}

As mentioned in Sec.~\ref{Sec:intro}, the CLAS high-statistics differential and total cross-section data for $\gamma p \to K^{*+}\Sigma^0$ \cite{Wei:2013} have so far only been analyzed by the work of Refs.~\cite{Kim:2013,Kim:20132}, where the resonances $N(2080)3/2^-$, $N(2090)1/2^-$, $N(2190)7/2^-$, $N(2200)5/2^-$, $\Delta(2150)1/2^-$, $\Delta(2200)7/2^-$, and $\Delta(2390)7/2^+$ have been introduced in addition to the $N$ and $\Delta$ exchanges, with the resonance electromagnetic couplings taken from a quark model estimation \cite{Capstick:1992}, the resonance hadronic couplings determined by resonance partial decay amplitudes calculated from Ref.~\cite{Capstick:1998}, and the signs of the resonance couplings being fitting parameters. It was claimed that the resonance contributions gave only negligible effects while the contributions of $t$-channel $K$ exchange and the $s$-channel $\Delta$ exchange are crucial for both $\gamma p \to K^{*+} \Sigma^0$ and $\gamma p \to K^{*0} \Sigma^+$ reaction processes.

As illustrated in Fig.~\ref{Fig:status}, although Refs.~\cite{Kim:2013,Kim:20132} provide a qualitative description of the CLAS high-statistics cross-section data, it is clearly seen that there is still some room for improvement. In the present work, we perform an analysis independent of Refs.~\cite{Kim:2013,Kim:20132} of the CLAS high-statistics cross-section data for $K^*\Sigma$ photoproduction reactions. We attempt to get a better description of the data, especially for the $\gamma p \to K^{*+} \Sigma^0$ reaction, which will allow for a more reliable understanding of the reaction mechanisms and the associated resonance parameters. We employ the effective Lagrangian approach. If we only consider the $t$-channel $K$, $\kappa$, $K^*$ exchanges, the $u$-channel $\Lambda$, $\Sigma$, $\Sigma^*$ exchanges, the $s$-channel $N$, $\Delta$ exchanges, and the generalized contact current as illustrated in Fig.~\ref{FIG:feymans} in constructing the reaction amplitudes, the fitting quality of the CLAS high-statistics differential cross-section data \cite{Wei:2013} will be worse than that illustrated by the dashed lines of Fig.~\ref{Fig:status}, which in our opinion cannot be treated as an acceptable description of the data. We then introduce the $s$-channel resonances as few as possible in order to achieve a satisfactory description of the data. We consider one by one the four-star and three-star near-threshold resonances with their masses, widths, and helicity amplitudes taken to be the averaged values advocated in the most recent version of RPP \cite{Patrignani:2016}. After many trials we found that, the data can be well described by including the $\Delta(1905)5/2^+$ resonance, a four-star resonance rated by RPP \cite{Patrignani:2016}. It is also found that introducing a three-star or four-star resonance other than $\Delta(1905)5/2^+$ will result in a much larger $\chi^2$ and obvious discrepancies in comparison with the data. Since so far we only have the cross-section data, we in the present work do not pursue an analysis with one one-star or two-star resonance whose masses, widths, and helicity amplitudes are unknown in RPP \cite{Patrignani:2016} or an analysis with one more resonance besides the $\Delta(1905)5/2^+$ in constructing the reaction amplitudes, as doing so will result in much more adjustable parameters that cannot be well constrained by the cross-section data alone. We postpone such attempts until the data for spin observables become also available. With this in mind, we conclude that the CLAS high-statistics differential and total cross-section data for $\gamma p \to K^{*+} \Sigma^0$ and $\gamma p \to K^{*0} \Sigma^+$ can be well described by including the $\Delta(1905)5/2^+$ resonance, and the corresponding results serve as an analysis independent of Refs.~\cite{Kim:2013,Kim:20132} for the CLAS data with less adjustable parameters and better fitting quality.

In our analysis, we use the averaged values in RPP for the mass $M_R$, width $\Gamma_R$, and helicity amplitudes $A_{1/2}$, $A_{3/2}$ for the resonance $\Delta(1905)5/2^+$:
\begin{align*}
M_R & \approx 1880 ~{\rm MeV}, &\quad A_{1/2} &\approx 0.022 ~{\rm GeV}^{-1/2}, \\
\Gamma_R &\approx 330 ~{\rm MeV}, &\quad A_{3/2} &\approx -0.045 ~{\rm GeV}^{-1/2}.
\end{align*}
The resonance hadronic couplings, $g_{R\Sigma K^*}^{(i)}$ $(i=1,2,3)$, are treated as fitting parameters. The other fitting parameters employed in the present work have already been introduced in Sec.~\ref{Sec:formalism}. The fitted values of all these adjustable parameters are listed in Table~\ref{Table:para}. There, the uncertainties in the resulting parameters are estimates arising from the uncertainties (error bars) associated with the fitted experimental differential cross-section data points.

\begin{table}[tb]
\caption{\label{Table:para} Model parameters. See Sec.~\ref{Sec:formalism} for their definitions.}
\begin{tabular*}{\columnwidth}{@{\extracolsep\fill}lr}
\hline\hline
$g^{(2)}_{\Sigma^{*+} \Sigma^{+} \gamma}/g^{(1)}_{\Sigma^{*+} \Sigma^{+} \gamma}$ & $3.10\pm0.33$ \\
$g^{(1)}_{\Sigma^{*0} \Sigma^{0} \gamma}$  & $0.74\pm0.25$ \\
$g^{(2)}_{\Sigma^{*0} \Sigma^{0} \gamma}$  &  $10.50\pm 2.62$  \\
$g_{\Delta \Sigma K^*}^{(1)}$ & $-8.84 \pm0.06$  \\
$g^{(1)}_{R \Sigma K*}$ & $3.70 \pm0.03$      \\
$g^{(2)}_{R \Sigma K* }$ & $-9.59\pm 0.32$     \\
$g^{(3)}_{R \Sigma K* }$ & $29.29\pm0.33   $    \\
$\Lambda_s$ [MeV]   & $1358\pm 2$    \\
$\Lambda_{\Sigma^*}$  [MeV]  &  $843 \pm 3$  \\
$\Lambda_{\Lambda}$ [MeV]  &  $797 \pm 8$  \\
$\Lambda_{\Sigma}$ [MeV]  &  $700 \pm 78$  \\
$\Lambda_{K,\kappa}$  [MeV]  &  $1197 \pm 56$  \\
$\Lambda_{K^{*}}$  [MeV]  &  $1233 \pm 26$  \\
\hline\hline
\end{tabular*}
\end{table}

\begin{figure*}[tbp]
\includegraphics[width=0.95\textwidth]{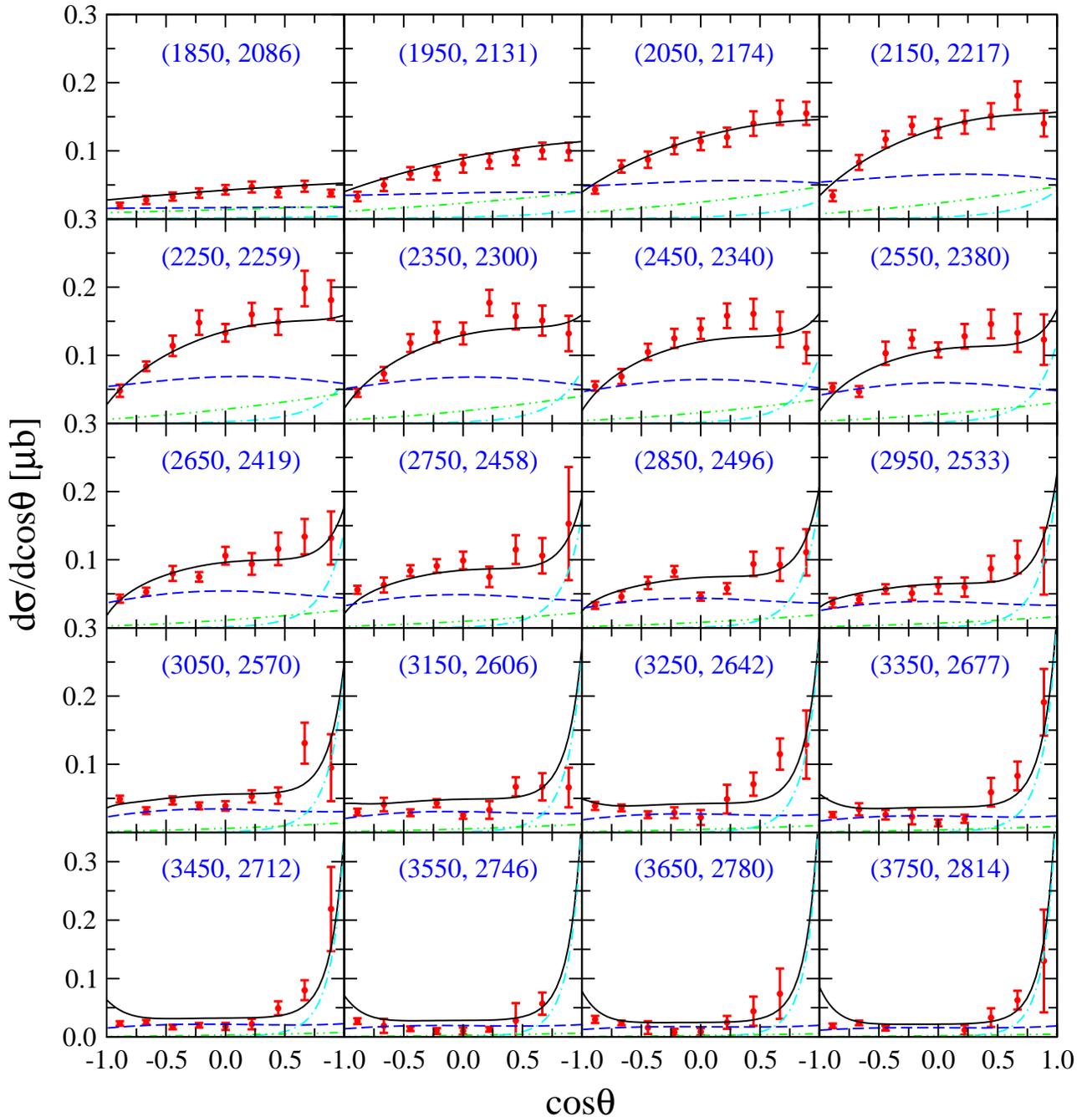}
\caption{Differential cross sections for $\gamma p \to K^{*+}\Sigma^0$ as a function of $\cos\theta$ (black solid lines). The scattered symbols denote the CLAS data in Ref.~\cite{Wei:2013}. The blue dashed, green dash-double-dotted, and cyan dash-dotted lines represent the individual contributions from the $\Delta(1905)5/2^+$, $\Delta$, and $K^*$ exchanges, respectively. The numbers in parentheses denote the centroid value of the photon laboratory incident energy (left number) and the corresponding total center-of-mass energy of the system (right number), in MeV.}
\label{fig:dif1}
\end{figure*}

\begin{figure*}[tbp]
\includegraphics[width=0.95\textwidth]{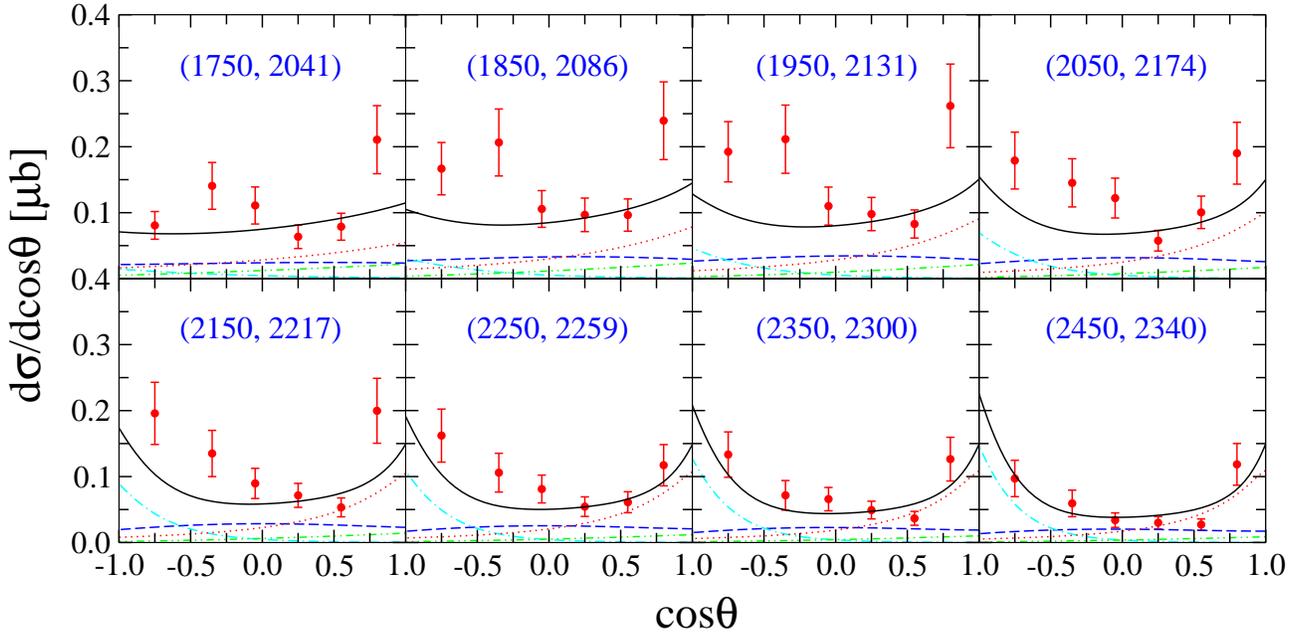}
\caption{Differential cross sections for $\gamma p \to K^{*0}\Sigma^+$ as a function of $\cos\theta$. Notations are the same as in Fig.~\ref{fig:dif1} except that now the cyan dash-dotted lines represent the contributions from the $u$-channel $\Sigma^*$ exchange, and the magenta dotted lines denote the contributions from the $t$-channel $K$ exchange. The scattered symbols denote the CLAS data in Ref.~\cite{Hleiqawi:2007ad}.}
\label{fig:dif2}
\end{figure*}

\begin{figure*}[tbp]
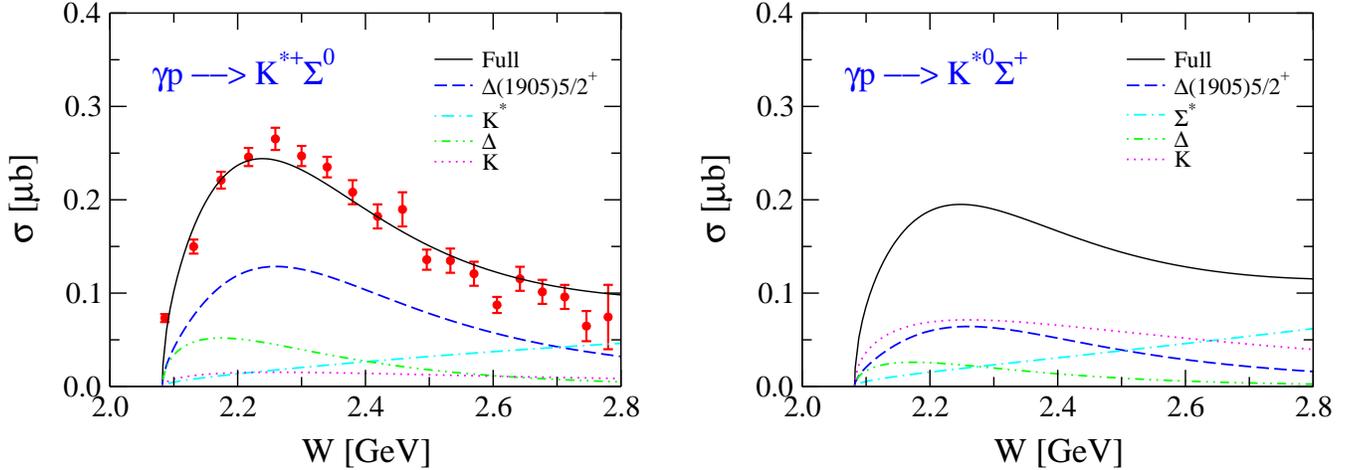

\vglue 0.1cm
\includegraphics[width=0.47\textwidth]{sig1}
\hglue 0.7cm
\includegraphics[width=0.47\textwidth]{sig2}
\caption{Total cross sections with dominant individual contributions for $\gamma p \to K^{*+}\Sigma^0$ and $\gamma p \to K^{*0}\Sigma^+$. The left graph is for $K^{*+}\Sigma^0$ channel and the right one corresponds to $K^{*0}\Sigma^+$ channel. The solid lines represent the full results. The blue dashed, green dash-double-dotted, and magenta dotted lines represent the individual contributions from the $\Delta(1905)5/2^+$, $\Delta$ and $K$ exchanges, respectively. The cyan dash-dotted line represents the $K^*$ exchange in the left graph and the $\Sigma^*$ exchange in the right one. The scattered symbols are data from CLAS Collaboration \cite{Wei:2013}.  }
\label{fig:total_cro_sec}
\end{figure*}

The results for differential cross sections of $\gamma p\to K^{*+}\Sigma^0$ and $\gamma p\to K^{*0}\Sigma^+$ corresponding to the model parameters listed in Table~\ref{Table:para} are shown in Fig.~\ref{fig:dif1} and Fig.~\ref{fig:dif2}, respectively. There, the black solid lines represent the full results. The blue dashed and green dash-double-dotted lines represent the individual contributions from the $\Delta(1905)5/2^+$ and $\Delta$ exchanges, respectively. The cyan dash-dotted lines represent the individual contributions from $K^*$ exchange for $\gamma p\to K^{*+}\Sigma^0$ and $\Sigma^*$ exchange for $\gamma p\to K^{*0}\Sigma^+$. The magenta dotted lines in Fig.~\ref{fig:dif2} denote the individual contributions from the $t$-channel $K$ exchange. The contributions from other terms are too small to be clearly seen with the scale used, and thus they are not plotted. The numbers in parentheses denote the centroid value of the photon laboratory incident energy (left number) and the corresponding total center-of-mass energy of the system (right number), in MeV. The statistical data binning for photon incident energy is $100$ MeV, whose effects for $\gamma p\to K^{*+}\Sigma^0$ at the center-of-mass energy $W=2086$ MeV, which is about 2 MeV higher than the $K^{*+}\Sigma^0$ threshold, have been approximated by an integral of the differential cross sections over the 100 MeV energy bin. At other energies, the binning effects have been tested to be tiny. One sees from Figs.~\ref{fig:dif1}--\ref{fig:dif2} that our overall description of the CLAS high-statistics angular distribution data is fairly satisfactory in the whole energy region considered, much better than the description from Refs.~\cite{Kim:2013,Kim:20132} (c.f. Fig.~\ref{Fig:status}). 

For $\gamma p\to K^{*+}\Sigma^0$, Fig.~\ref{fig:dif1} shows that the $\Delta(1905)5/2^+$ exchange (blue dashed lines) provides dominate contributions to the cross sections at low energies. The contributions from the $\Delta$ exchange (green dash-double-dotted lines) are considerable in the region near the $K^{*+}\Sigma^0$ threshold. At high energies, the differential cross sections are forward-peaked and dominated by the $t$-channel $K^*$ exchange (cyan dash-dotted lines). Note that our results are quite different from those from Refs.~\cite{Kim:2013,Kim:20132}, where it was found that the angular distributions for $\gamma p\to K^{*+}\Sigma^0$ are nearly described by the $\Delta$ exchange alone in the whole energy region, while the contributions form all other resonances are negligible, and the contributions from $K^*$ exchange are also rather small. The differences can be roughly understood from the following analysis. In our present work, the cutoff for $K^*$ exchange is a fit parameter, and the fitted value is $\Lambda_{K^*} \approx 1.2$ GeV as listed in Table~\ref{Table:para}, while in Refs.~\cite{Kim:2013,Kim:20132}, $\Lambda_{K^*}$ is chosen to be 0.8 GeV, much smaller than our fitted value. This explains why Refs.~\cite{Kim:2013,Kim:20132} have smaller contributions from $K^*$ exchange than ours. Note that even our results are in good agreement with the data, the contributions from $K^*$ exchange needs to be further constrained by the data at very forward angles at high energies, which are sparse at the moment. For $\Delta$ exchange, both the coupling constant and the cutoff from our work are smaller than those used in Refs.~\cite{Kim:2013,Kim:20132}, leading to smaller $\Delta$ contributions in our work. For resonances, Refs.~\cite{Kim:2013,Kim:20132} include the $N(2080)3/2^-$, $N(2090)1/2^-$, $N(2190)7/2^-$, $N(2200)5/2^-$, $\Delta(2150)1/2^-$, $\Delta(2200)7/2^-$, and $\Delta(2390)7/2^+$ with the electromagnetic couplings taken from a quark model estimation \cite{Capstick:1992}, the hadronic couplings calculated from Ref.~\cite{Capstick:1998}, and the signs of the resonance couplings being fitting parameters. In our work, we consider only $\Delta(1905)5/2^+$ with its mass, width, helicity amplitudes taken from the most recent edition of RPP \cite{Patrignani:2016} and its hadronic couplings being fitting parameters. One sees that the data are described quite well in our present work.

For $\gamma p\to K^{*0}\Sigma^+$, Fig.~\ref{fig:dif2} shows that the angular dependences are  dominated by the $t$-channel $K$ exchange (magenta dotted lines) at forward angles and the $u$-channel $\Sigma^*$ exchange (cyan dash-dotted lines) at backward angles. The $s$-channel $\Delta(1905)5/2^+$ exchange (blue dashed lines) makes considerable contributions at low energies, and the $s$-channel $\Delta$ exchange (green dash-double-dotted lines) gives small but non-negligible contributions near-threshold. These observations are also different from Refs.~\cite{Kim:2013,Kim:20132}, where it was claimed that the $s$-channel $\Delta$ exchange provides dominant contributions while the contributions from all other resonances are negligible. The main reason for the differences is the same as mentioned above for $\gamma p\to K^{*+}\Sigma^0$ reaction. Note that for $\gamma p\to K^{*0}\Sigma^+$ reaction, the results from both our work and Refs.~\cite{Kim:2013,Kim:20132} are in qualitative agreement with the data, which have larger error bars than those for $\gamma p\to K^{*+}\Sigma^0$. 

In Ref.~\cite{Oh:2006in}, based on an investigation of the very preliminary data for $\gamma p \to K^{*+} \Lambda$ and $\gamma p \to K^{*0}\Sigma^+$, it is claimed that the $t$-channel $\kappa$ exchange provides significant contributions to the reaction $\gamma p \to K^{*0}\Sigma^+$. In our present work, the contribution from the $t$-channel $\kappa$ exchange is found to be negligible. The same observation has also been found in Refs.~\cite{Kim:2013,Kim:20132}.

\begin{figure*}[tbp]
\includegraphics[width=0.75\textwidth]{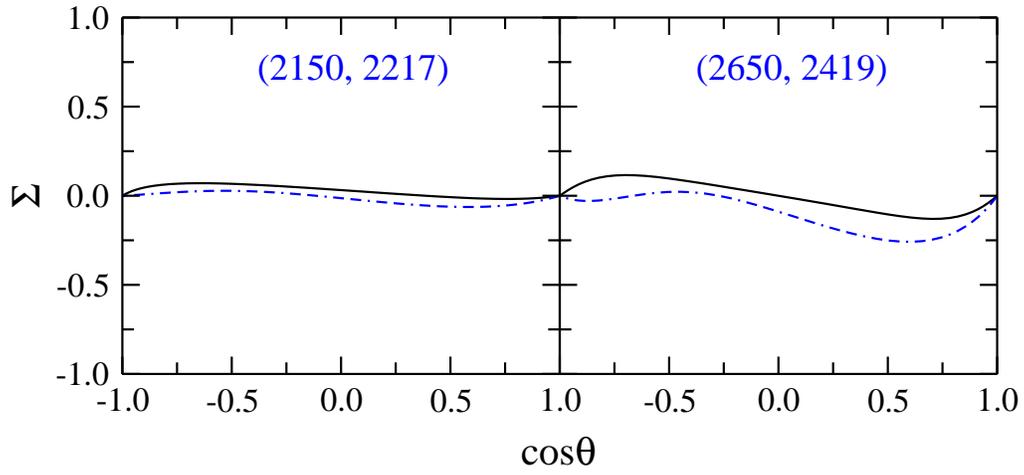}
\caption{Photon beam asymmetries as a function of $\cos\theta$.
The black solid lines represent the results for $\gamma p \to K^{*+}\Sigma^0$ while the blue dashed lines correspond to the results for $\gamma p\to K^{*0}\Sigma^+$. The numbers in parentheses denote the photon laboratory incident energy (left number) and the total center-of-mass energy of the system (right number), in MeV.}
\label{fig:beam_asy}
\end{figure*}

\begin{figure*}[tbp]
\vglue 0.3cm
\includegraphics[width=0.75\textwidth]{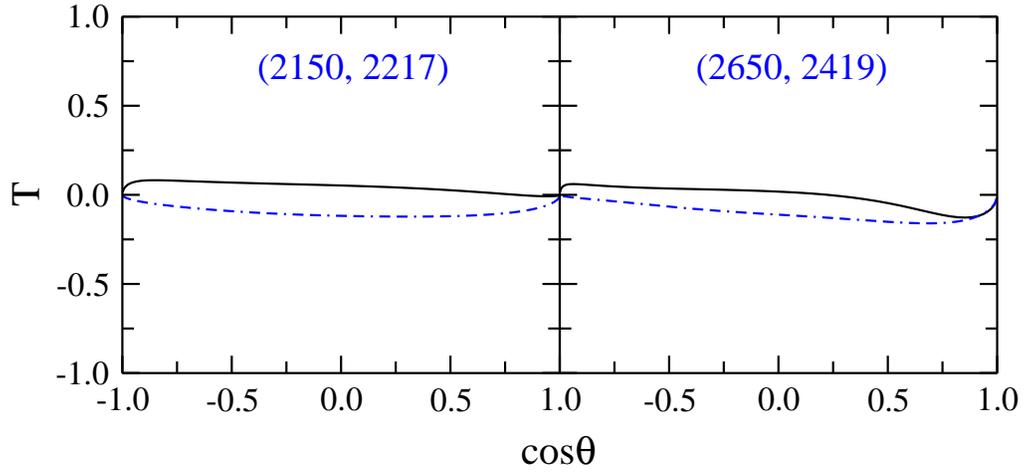}
\caption{Same as in Fig.~\ref{fig:beam_asy} for target nucleon asymmetries.}
\label{fig:target_asy}
\end{figure*}

\begin{figure*}[tbp]
\vglue 0.3cm
\includegraphics[width=0.75\textwidth]{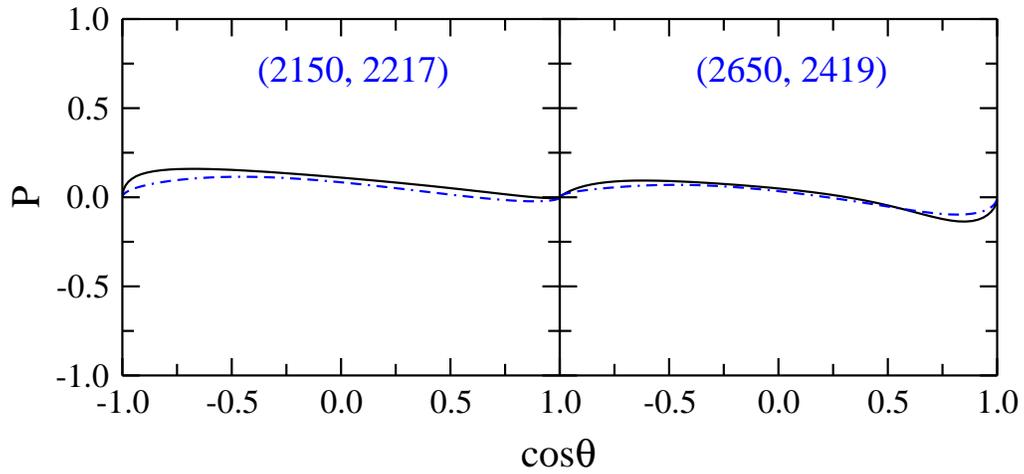}
\caption{Same as in Fig.~\ref{fig:beam_asy} for recoil $\Sigma$ baryon asymmetries.}
\label{fig:recoil_asy}
\end{figure*}

Figure~\ref{fig:total_cro_sec} shows our predicted total cross sections (black solid lines) together with individual contributions from the $\Delta(1905)5/2^+$ exchange (blue dashed lines), the $\Delta$ exchange (green dash-double-dotted lines),  and the $K$ exchange (magenta dotted lines) for $\gamma p\to K^{*+}\Sigma^0$ (left graph) and $\gamma p\to K^{*0}\Sigma^+$ (right graph). The cyan dash-dotted lines represent the $K^*$ exchange in the left graph and the $\Sigma^*$ exchange in the right one. The contributions from other terms are not plotted since they are too small to be clearly seen with the scale used. Note that the total cross-section data are not included in our fit procedure. One sees that our predictions are in agreement with the data in the full energy region considered. It is more clear to see in Fig.~\ref{fig:total_cro_sec} the importance of the resonance $\Delta(1905)5/2^+$ exchange. For $\gamma p\to K^{*+}\Sigma^0$, it causes the broad bump exhibited by the total cross sections. For $\gamma p\to K^{*0}\Sigma^+$, the $K$ exchange is as important as the $\Delta(1905)5/2^+$ exchange, and they two dominate the broad bump exhibited in the total cross sections. Note that the $K$ exchange has little contribution in the other channel due to the isospin factor together with the smaller electromagnetic coupling constant. The $\Delta$ exchange is seen to provide considerable contributions to both reactions. At high energies, the $K^*$ exchange becomes important for $\gamma p\to K^{*+}\Sigma^0$ due to its dominant contribution at forward angles as shown in Fig.~\ref{fig:dif1}, and the $\Sigma^*$ exchange becomes important for $\gamma p\to K^{*+}\Sigma^0$ due to its large contribution at backward angles as shown in Fig.~\ref{fig:dif2}.

In Figs.~\ref{fig:beam_asy}--\ref{fig:recoil_asy}, we show the predictions of the photon beam asymmetry ($\Sigma$), target nucleon asymmetry ($T$), and the recoil $\Sigma$ baryon asymmetry ($P$) from our present model. There, the solid and dash-dotted lines represent the corresponding results for $\gamma p\to K^{*+}\Sigma^0$ and $\gamma p\to K^{*0}\Sigma^+$, respectively. We hope that these spin observables can be measured in experiments in the near future, which can help to further constrain the model and thus result in a better understanding of the reaction mechanisms and the associated resonance contents and parameters.

\section{Summary and conclusion}  \label{sec:summary}

In the present work, we employ an effective Lagrangian approach at the tree-level Born approximation to analyze the available data for the two-channel photoproduction reactions, $\gamma p \to K^{*+} \Sigma^0$ and $\gamma p \to K^{*0} \Sigma^+$. It is found that if we only consider the contributions from the $t$-channel $K$, $\kappa$, $K^*$ exchanges, the $s$-channel $N$, $\Delta$ exchanges, the $u$-channel $\Lambda$, $\Sigma$, $\Sigma^*$ exchanges, and the generalized contact current, the data cannot be well described. We then try to include as few as possible the nucleon resonances in constructing the reaction amplitudes. We check one by one the four-star and three-star near-threshold resonances with their masses, widths, and helicity amplitudes taken to be the averaged values advocated in the most recent RPP \cite{Patrignani:2016}, and find that the inclusion of the four-star resonance $\Delta(1905)5/2^+$ results in a satisfactory description of the differential and total cross-section data reported by the CLAS Collaboration, while the inclusion of other four-star or three-star resonance leads to a much bigger $\chi^2$ and obvious discrepancies in comparison with the data. We do not pursue an analysis with one one-star or two-star resonance whose mass, width, and helicity amplitudes are unknown in RPP \cite{Patrignani:2016} or an analysis with one more resonance besides $\Delta(1905)5/2^+$ in the present work, since doing so will give rise to much more adjustable parameters, which cannot be well determined by the available cross-section data. We postpone such attempts until the data for spin observables become also available. 

The present work provides an analysis of the CLAS high-precision cross-section data for $\gamma p \to K^{*+} \Sigma^0$ and $\gamma p \to K^{*0} \Sigma^+$ independent of Refs.~\cite{Kim:2013,Kim:20132}. It should be mentioned that the reaction mechanisms in these two models are quite different, although the present paper has less adjustable parameters and a better fitting quality. In Refs.~\cite{Kim:2013,Kim:20132}, the contributions from the resonances $N(2080)3/2^-$, $N(2090)1/2^-$, $N(2190)7/2^-$, $N(2200)5/2^-$, $\Delta(2150)1/2^-$, $\Delta(2200)7/2^-$, and $\Delta(2390)7/2^+$ are considered, and they all are found to be negligible compared with the dominant $s$-channel $\Delta$ exchange and $t$-channel $K$ exchange. In the present work, it shows that the $s$-channel $\Delta(1905)5/2^+$ resonance contributes significantly to both $\gamma p \to K^{*+} \Sigma^0$ and $\gamma p \to K^{*0} \Sigma^+$ reactions. The $s$-channel $\Delta$ exchange provides considerable but less important contributions near $K^*\Sigma$ threshold in both reactions. At high energies, the cross sections of the reaction $\gamma p \to K^{*+} \Sigma^0$ are dominated by the $t$-channel $K^*$ exchange, which causes a peak at forward angles. The cross sections of $\gamma p \to K^{*0} \Sigma^+$ at high energies are dominated by the $u$-channel $\Sigma^*$ exchange at backward angles and the $t$-channel $K$ exchange at forward angles.

The predictions of the photon beam asymmetry ($\Sigma$), target nucleon asymmetry ($T$), and the recoil $\Sigma$ baryon asymmetry ($P$) from the present model are also presented for both $\gamma p\to K^{*+}\Sigma^0$ and $\gamma p\to K^{*0}\Sigma^+$ reactions. The shape of all of them are quite different from those predicted in Refs.~\cite{Kim:2013,Kim:20132}. High-statistic data on those spin observables are expected to further constrain the model and help one get a better understanding of the reaction mechanisms.

\begin{acknowledgments}
This work is partially supported by the National Natural Science Foundation of China under Grants No.~11475181 and No.~11635009, the Youth Innovation Promotion Association of Chinese Academy of Sciences under Grant No.~2015358, and the Key Research Program of Frontier Sciences of Chinese Academy of Sciences under Grant No. Y7292610K1.
\end{acknowledgments}

\end{document}